\documentclass{amsart}
\usepackage{amsmath}
\usepackage{amssymb}
\usepackage{url}

\usepackage{yhmath}

\usepackage{srctex} 

\newcommand{\cl}[2]{\ensuremath{\mathit{Cl}_{#1,#2}}} 

\newcommand{\reverse}[1]{\widetilde{#1}}
\newcommand{\gradeinverse}[1]{\wideparen{#1}}
\newcommand{\cliffordconjugate}[1]{\wideparen{\widetilde{#1}}}

\newcommand{\e}[1]{\mathbf{e}_{#1}} 

\def\m#1{\mathsf{#1}} 

\begin{document}


\title[Inverse of  multivector: Beyond p+q=5]
 {Inverse of  multivector: Beyond p+q=5\\ threshold}

\author{A.~Acus}
\address{Institute of Theoretical Physics and Astronomy,
Vilnius University, Saul{\.e}tekio 3, LT-10257 Vilnius,
Lithuania} \email{arturas.acus@tfai.vu.lt}

\author{A.~Dargys}

\address{%
Center for Physical Sciences and Technology, Semiconductor
Physics Institute, Saul{\.e}tekio 3, LT-10257 Vilnius,
Lithuania}

\email{adolfas.dargys@ftmc.lt}


\date{June, 2017}

\begin{abstract}
The algorithm of finding inverse multivector (MV) numerically and
symbolically is of paramount importance in the applied Clifford
geometric algebra (GA) \cl{p}{q}. The first general MV inversion
algorithm was based on matrix representation of MV. The complexity
of calculations and size of the answer in a symbolic form
grow exponentially with the GA dimension $n=p+q$. The breakthrough
occurred when D.~Lundholm and then P.~Dadbeh found compact inverse
formulas up to dimension $n\le5$. The formulas were constructed in
a form of Clifford product of initial MV and its carefully chosen
grade-negation counterparts. In this report we show that the
grade-negation self-product method can be extended beyond $n=5$
threshold if, in addition, properly constructed linear
combinations of such MV products are used. In particular, we
present compact explicit MV inverse formulas for algebras of
vector space dimension $n=6$ and show that they embrace all lower
dimensional cases as well. For readers convenience, we have also
given various MV formulas in a form of grade negations when
$n\le5$.
\end{abstract}

\maketitle

\section{Introduction\label{sec:1}}

The knowledge of how to find inverse multivector (MV) in the
Clifford algebra in a symbolic and coordinate-free form is very
important both from practical computational and purely theoretical
point of views. A universal formula for inverse MV would allow to
write down a fast and general algorithm for all occasions rather
then to resort to either specific  symbolic or numerical cases.
The inverse of MV can be used to find explicit solutions of
algebraic GA equations with all the ensuing consequences and
applications. A closely related problem is the normalization of
spinors~\cite{Vaz2016,Lundholm09}. Invertibility also serves as an
important criterion on deciding whether homogeneous versors are
the blades~\cite{Bouma2001}, which are essential in numerous
geometric constructions.

The first attempts of inversion of some specific forms of MVs can
be traced back to papers~\cite{Semenov1991,
Semenov1993A,Semenov1993B}. However it was not until 2002 when
J.P.~Fletcher \cite{Fletcher2002} has suggested some general MV
inverse formulas for low dimensional algebras. His algorithm was
based on decomposition of MV into matrix basis elements, and, as a
consequence, resulted into large, inconvenient  and signature
dependent formulas the size of which grows exponentially with
algebra dimension.

The breakthrough occurred after D.~Lundholm~\cite{Lundholm06} has
presented explicit expressions for $n\le5$ (determinant) norms and
later P.~Dadbeh~\cite{Dadbeh2011} introduced grade-negation
operation what has allowed to write down compact and explicit
formulas for MV inverse in a coordinate-free form. The heart of
algorithm~\cite{Dadbeh2011} is the geometric product of initial MV
and its carefully chosen grade-negated counterpart(s) that after
few iterations eventually yields a scalar. As a matter of fact,
the product may be related with the determinant of a matrix
representation of general MV, from which the inverse multivector
can be easily extracted  by simply removing the initial MV which
is always positioned in either left-most or right-most side of the
product. Using the described method P.~Dadbeh was able to find
explicit inverses for a general MV up to dimension $n\le 5$. It is
important to stress that the obtained formulas are determined by
vector space dimension $n=p+q$ only and are independent of a
particular GA signature $(p,q)$. The same formulas were also
obtained by other authors using different methods (see, for
example,~\cite{Shirokov2012,Hitzer2016,Suzuki2016}). When $n\le
5$, detailed mathematical proofs are given in \cite{Hitzer2016}.
If general multivector $\m{A}$ is given in expanded form in some
orthogonal basis with symbolic coefficients, then the verification
of the algorithm  can be easily done by a direct substitution of
symbolic MV into formula and explicitly computing the inverse
$\m{A}^{-1}$, and finally checking that the property
$\m{A}\m{A}^{-1}=\m{A}^{-1}\m{A}=1$ is satisfied. For $n>4$,
however, calculation of explicit inverse in symbolic form is time
consuming and results in lengthy expressions for  coefficients at
basis elements of $\m{A}^{-1}$. Nonetheless, such calculations, in
fact, have status of ``computer-assisted proof''. Similar formulas
for the (determinant) norms of MVs when $n\le 5$ were also given
in~\cite{Lundholm09}. Analysis of general structure of such
formulas was presented in~\cite{Suzuki2016}. However, until now
any attempts to step across the threshold $p+q=5$ were
unsuccessful although there is a need for such formulas in
practice.

In this report we show that the grade-negation method can be
extended beyond $n=5$ threshold if, in addition, properly
constructed linear combinations of grade-negated MVs are
introduced. In particular, we write down explicit MV inverse
formulas for algebras with vector space dimension $n=6$ having all
possible $(p,q)$ signatures. We also provide alternative formulas
for even MVs and $n=5$ case. In Sec.~\ref{sec:2} the
grade-negation method is shortly reviewed and required notation
and terminology is introduced. In Sec.~\ref{sec:3} the inverse
even MVs that follow from higher grade inverse MVs are considered.
Finally, in Sec.~\ref{sec:4} a general algorithm for construction
of inverse MV at $n=6$ is briefly discussed and the obtained
coordinate-free formulas are presented in a form of tables.

\section{Grade-negated self-product and inverse of a general MV in $n\le5$ case\label{sec:2}}

Following~\cite{Dadbeh2011} we first introduce a grade-negated
self-product that is defined via \textit{grade-$r$ negation
operation}.  Applied to the multivector $\m{A}$ this operation
does what it says, i.~e., it changes the sign of grade-$r$ part of
$\m{A}$.  Such grade-$r$ negated MV will be denoted as
$\m{A}_{\bar r}$, with the bar over index designating which of the
grades have opposite signs.  Formally the grade-$r$ negated MV can
be expressed as $\m{A}_{\bar r}=\m{A}-2\langle \m{A}\rangle_r$, or
$\m{A}_{\bar r,\bar s}=\m{A}-2\langle \m{A}\rangle_r-2\langle
\m{A}\rangle_s$ for a double negation, where $\langle
\m{A}\rangle_r$ denotes grade-$r$ projection of multivector
$\m{A}$. In particular, we have $(\langle\m{A}\rangle_r)_{\bar
r}=-\langle\m{A}\rangle_r$.

The following properties are evident  from the definition of
grade-negation operation: $(\m{A}_{\bar p})_{\bar r}=\m{A}_{{\bar
p},{\bar r}}=\m{A}_{{\bar r},{\bar p}}$, $\m{A}_{{\bar r},{\bar
r}}=\m{A}$, $(\m{A}+\m{B})_{\bar r}=\m{A}_{\bar r}+\m{B}_{\bar
r}$. However, $(\m{AB})_{\bar r}\ne \m{A}_{\bar r}\m{B}_{\bar r}$.
If $\m{A}$ does not contain grade-$r$ elements then negation
returns the same MV, $\m{A}_{\bar r}=\m{A}$. Commutator with
grade-negated MV then can be expressed as $\m{A} \m{A}_{\bar
r}-\m{A}_{\bar r}\m{A}=2(\langle \m{A}\rangle_r\m{A}-\m{A}\langle
\m{A}\rangle_r)$. Multivector $\m{A}$ commutes with scalar
negated $\m{A}$, i.~e., $[\m{A},\m{A}_{\bar 0}]=0$ and, as a
consequence, with $\m{A}_{\bar i,\cdots,\bar j}$, where all grades
$i\neq0,j\neq 0,\cdots$ of $\m{A}$ (except scalar) are negated.

The standard involutions such as MV reversion $\reverse{\m{A}}$,
grade inversion $\gradeinverse{\m{A}}$, and Clifford conjugate
$\cliffordconjugate{\m{A}}=\reverse{\gradeinverse{\m{A}}}$ in
terms of negation can be written as $\reverse{\m{A}}=\m{A}_{{\bar
2},{\bar 3},{\bar 6},{\bar 7},{\bar {10}},{\bar {11}}\dots}$,
$\gradeinverse{\m{A}}=\m{A}_{{\bar 1},{\bar 3},{\bar 5},{\bar
7},{\bar {9}},{\bar {11}}\dots}$ and
$\cliffordconjugate{\m{A}}=\m{A}_{{\bar 1},{\bar 2},{\bar 5},{\bar
6},{\bar {9}},{\bar {10}}\dots}$, respectively. For finite
algebras the index series should  be chopped off when negation
index becomes larger than that of the pseudoscalar. It is worth
mentioning  that the grades $0,4,8,12,\dots$ are absent in the
above list of standard GA involutions. From this follows that the
inverse of a general MV cannot be expressed using just standard
involutions, or their combinations. As we shall see this
observation also is directly related to the MV inversion problem
for $n\ge 6$.

By grade-negated self-product we identify the geometric product of
general MV $\m{A}$ with any number of its grade-negated
counterparts, for example, geometric product $\m{A} \m{A}_{\bar r}
\m{A}_{\bar k,\bar l,\bar t} \m{A}_{\bar s, \bar t}\cdots$ is the
left grade-negated self-product and $\m{A}_{\bar r} \m{A}_{\bar
k,\bar l,\bar t} \m{A}_{\bar s, \bar t}\cdots\m{A}$ is the right
grade-negated self-product. These self-products, where the initial
MV  stands in the left-most or right-most position, will be of
special importance in finding the inverse MV $\m{A}^{-1}$.

The explicit inverse formulas which we will construct rely on our
ability to get a real scalar using just grade-negated
self-products (or, as we shall see, some linear combinations of
them for $n>5$), where initial MV $\m{A}$ is mandatory and is
located either in the left-most or right-most position. We shall
call the real scalar
\begin{equation}\label{discriminant}
s_m=\overbrace{\m{A}\m{A}_{\bar r,\bar t,\ldots}\m{A}_{\bar
s,\ldots}\cdots}^m = \m{A}\, f\bigl(\m{A}\bigr)= f\bigl(\m{A}\bigr)\,\m{A}
\end{equation}
obtained in this way the determinant norm\footnote{In literature,
the GA terminology on MV norms  is quite confusing. The
book~\cite{Hestenes1987} and most of introductory GA lecture
courses, for example~\cite{Chisolm2012}, typically define the MV
norm $|\m{A}|^2$ as a scalar part of product
$\left\langle\m{A}\reverse{\m{A}}\right\rangle_0$, which can be
negative.  (A better name would be the pseudonorm). If sign is
positive this kind of norm is also called the magnitude. The same
term sometimes is used to define a positive square root of
absolute value of $\sqrt{\mathrm{abs}(|\m{A}|^2)}$. On the other
hand, in~\cite{Lundholm06} a different scalar (named MV norm)
is defined which coincides with our determinant norm $s_m$ for $n\le 5$, whereas in~\cite{Suzuki2016} a
similar construction is called ``the discriminant''.
In~\cite{Dadbeh2011,Shirokov2012} this construction has been named
``the determinant'', because a determinant of matrix
representation of $\m{A}$ always coincides with the scalar. Other
authors~\cite{Hitzer2016} bypass naming confusion calling it just
``a real scalar''. In order to maintain the analogy with MV norm
defined in~\cite{Lundholm06} and to distinguish it from the norm
definition of~\cite{Hestenes1987}, we will use prefixed form
``determinant (pseudo)norm'' or shorter form ``determinant norm'',
or  just a ``determinant'' if it is clear from the context
what is meant. In general, the last term  should not
be confused with a determinant of matrix that  represents 
the MV~$\m{A}$. It should be noted, however, that the term
``determinant'' in GA is usually addressed in the context of
linear transformations~(see, for example, \cite{DoranLasenby2003},
p.~108, the subsection ``The determinant''), the definition and
meaning of which is related to the exterior product of basis
vectors and has nothing to do neither with the above
determinant norm nor with a determinant of matrix that
represents $\m{A}$.}.  If we succeed in constructing such a scalar
$s_m$  then it is easy to see that inverse MV formula for
$\m{A}$ (correspondingly, for $\reverse{\m{A}}$) can be obtained
simply by removing either extreme left initial MV $\m{A}$  or
extreme right initial MV $\m{A}$ from the scalar scalar $s_m$
in~\eqref{discriminant} (correspondingly, from the
$s_m^{\prime}=\cdots \reverse{\m{A}}_{\bar
s,\ldots}\reverse{\m{A}}_{\bar r,\bar t \ldots}\reverse{\m{A}}$),
and then dividing the result by $s_m$ or
$s_m^{\prime}=\reverse{\m{A}}\,
\reverse{f}\bigl(\reverse{\m{A}}\bigr)=
\reverse{f}\bigl(\reverse{\m{A}}\bigr)\,\reverse{\m{A}}$. The
inverse formulas for a MV $\m{A}$, thus, become
\begin{equation}\label{inverse}
\m{A}^{-1}=\frac{\m{A}_{\bar r,\bar t,\ldots}\m{A}_{\bar
s,\ldots}\cdots}{s_m}=\frac{f\bigl(\m{A}\bigr)}{s_m},\quad\reverse{\m{A}^{-1}}=\frac{\cdots \reverse{\m{A}}_{\bar s,\ldots}\reverse{\m{A}}_{\bar r,\bar t \ldots}}{s_m^{\prime}}=\frac{\reverse{f}\bigl(\reverse{\m{A}}\bigr)}{s_m^{\prime}}.
\end{equation}
All known cases for $n\le5$ (for detailed proofs
see~\cite{Hitzer2016}) show that inverse formulas
in~\eqref{inverse} can be applied  to  arbitrary signature algebra
at a fixed vector space dimension $n$. Furthermore, it is easy to
check explicitly, for example, that the inverse formula for $n=5$
also yields inverses for $n=4,3,2,1$ as well, i.~e., each of
formula for larger $n$ {\it automatically contains inverses of all
lower algebras $m\le n$ with all possible signatures}. Therefore,
we {\it conjecture} that explicit formulas for determinant
norm~\eqref{discriminant} are signature independent\footnote{The informal explanation is
very simple. Geometric product of two vectors
splits into antisymmetric and symmetric parts: $a b=a\wedge b +
a\cdot b$. Because the commutators vanish, $[e_i,e_i]=0$ for
$i=1,2,...$, only the diagonal part (i.~e. signature) is not fixed
in the totaly antisymmetric expression (whatever that might mean),
when the geometric product of vectors is extended to the whole
algebra (whatever that might mean). Since the result (the determinant
norm) is a real scalar obtained by the same fixed antisymmetrization construction (i.~e. the formula), it can only be a function of signature.}, and that,
when restricted to lower dimension vector spaces, they yield
determinant norms of lower dimensional vector spaces, generally
raised in some power, which can be easily determined from the
dimension of matrix representation (see \textit{Example~5} below).
The known cases also suggest that formulas~\eqref{inverse} always
ensure that the inverse commutes with all three main involutions:
reversion, grade inversion and Clifford conjugate, i.~e.
$\reverse{\m{A}^{-1}}=(\reverse{\m{A}})^{-1}$,
$\gradeinverse{\m{A}^{-1}}=(\gradeinverse{\m{A}})^{-1}$ and
$\cliffordconjugate{\m{A}^{-1}}=(\cliffordconjugate{\m{A}})^{-1}$.
This is not generally true for an arbitrary involution, for
example, for an arbitrary combination of grade negations:
$(\m{A}^{-1})_{\bar i,\ldots}\neq(\m{A}_{\bar i,\ldots})^{-1}$. In
all known cases we also have $s_m=s_m^{\prime}$.

It is well known that Clifford algebras are isomorphic to algebras
of square matrices, the left and right inverses of which coincide.
The Clifford MV, therefore, has only one inverse which, depending
on our needs, can be written using either of the two determinant
norm forms~\eqref{discriminant}. As a result we have that
$\m{A}\m{A}^{-1}=\m{A}^{-1}\m{A}=1$ which can also serve as a test
of correctness of GA inverse algorithm.

{\bf Two-dimensional quadratic space.} Since the one-dimensional
case is trivial we start with the algebras $Cl_{2,0}$, $Cl_{1,1}$
and $Cl_{0,2}$. For $n=2$ let it be $Cl_{2,0}$. Writing $\m{A} =
\sum_{J=0}^{2^n-1} a_J\e{J}$, where the multi-index $J$ covers all
orthonormal base elements arranged in the increasing order of the
degrees followed by the lexicographic order, i.~e., the lowest
grade elements appear first\footnote{As known, the summation is
orderless with respect to base elements.  Nevertheless, it is
convenient in advance to settle some order which is
required if we want to enumerate coefficients $a_i$ in front of
base elements in a unique way.} while elements of the same grade
are ordered lexicographically. In the sum  the multi-index takes
the following explicit values $J=[\{\},\{1\},\{2\},\{1,2\}]$,
which illustrate inverse degree lexicographic
ordering~\cite{Cox2007}.  We can check that self-product $\m{A}
\m{A}_{\bar 1,\bar 2}$ immediately gives the required scalar
$s_2=a_{\{\}}^2 - a_{1}^2 - a_{2}^2 + a_{1,2}^2$. In the standard
notation it is more convenient to rewrite the coefficient indices
as in $s_2=a_{0}^2 - a_{1}^2 - a_{2}^2 + a_{3}^2$, where the
scalar coefficient was indexed by zero. The indices of
coefficients for vector components run from $1$ to $n$, while
numerical indices for higher grades increase monotonically up to
$2^n$. In general, when programming, for a coefficient in front of
grade-$r$ base it is convenient to start the element index
enumeration by $\sum_{k=0}^{r-1} (\begin{smallmatrix}n\\
k\end{smallmatrix})$ and to end by
$\Bigl(\sum_{k=0}^{r-1} (\begin{smallmatrix}n\\
k\end{smallmatrix})\Bigr)+(\begin{smallmatrix}n\\
r\end{smallmatrix})-1$.  Here $(\begin{smallmatrix}n\\
k\end{smallmatrix})$ denotes the binomial coefficient. For
example, for \cl{3}{0}, which has three vectors and three
bivectors, these expressions give $1$ and $3$ for first/last
vector indices, and $4$ and $6$ for first/last bivector indices.
Taking into account that binomial with negative index vanishes,
this convention enumerates all coefficients of MV and allows an
easy transition to more standard notation in a consistent way, for
example,  for initial MV we write $\m{A}=a_0 + a_1\e{1} + a_2\e{2}
+ a_3\e{12}$ and  for grade-negated MV $\m{A}_{\bar 1,\bar 2}=a_0
- a_1\e{1}- a_2\e{2}- a_3\e{12}$.  This notation will be used
throughout the paper. Note, that indices of base elements will
never be renamed and always be referred to as multi-indices.
Before going to higher dimensional algebras a  few comments
 are in place
here. The comments are of common character and can be easily
generalized to higher dimensional algebras.

 (1) The explicit form of determinant $s_m$ comprises all coefficients $a_i$ of
a multivector~$\m{A}$.

 (2) The condition for a MV to have inverse is determined by
determinant which must  be nonzero, $s_m\ne 0$.  If some of
intermediate results in $s_m$ after negation reduce to zero the
entire determinant  turns out to zero automatically. This explicit
statement was given  in order to resolve indeterminate cases like
zero division by zero.

 (3) Because reversion operation leaves the scalar (determinant norm) invariant,
the above formula for $n=2$ can be rewritten as $s_2=\m{A}
\m{A}_{\bar 1,\bar 2}=\m{A}_{\bar 1,\bar 2}\m{A}=\reverse{\m{A}}_{\bar 1,\bar
2}\reverse{\m{A}}=\reverse{\m{A}}\reverse{\m{A}}_{\bar 1,\bar
2}$ for an arbitrary multivector $\m{A}$. The right
hand side of equality, therefore, can be understood as the
determinant norm of a multivector $\reverse{\m{A}}$ written in the
right hand side form (where $\reverse{\m{A}}$ now stands in the
right most position). Due to this arbitrariness, inversions can
always be written in two forms: $\m{A}^{-1}=\m{A}_{\bar 1,\bar
2}/s_2= \m{A}_{\bar 1,\bar   2}/(\m{A}\m{A}_{\bar 1,\bar 2})= \m{A}_{\bar 1,\bar   2}/(\m{A}_{\bar 1,\bar 2}\m{A})$ and, correspondingly,
$\reverse{\m{A}^{-1}}=\reverse{\m{A}}_{\bar 1,\bar
2}/(\reverse{\m{A}}_{\bar 1,\bar 2}\reverse{\m{A}})=\reverse{\m{A}}_{\bar 1,\bar
2}/(\reverse{\m{A}}\reverse{\m{A}}_{\bar 1,\bar 2})=\m{A}_{\bar 1,\bar
3}/(\m{A}_{\bar 2,\bar 3}\m{A}_{\bar 1,\bar 3})=\m{A}_{\bar 1,\bar
3}/(\m{A}_{\bar 1,\bar 3}\m{A}_{\bar 2,\bar 3})=\reverse{\m{A}}^{-1}$, where we explicitly had used $\reverse{\m{A}}=\m{A}_{\bar 2,\bar 3}$. Determinant
norms of $\m{A}$ and $\reverse{\m{A}}$ are
equal~\cite{Shirokov2012}, because the matrix representation of
the reversed MV yields the same determinant. Construction of
matrix operation itself, which corresponds to MV $\reverse{\m{A}}$
reversion for any signature is described in~\cite{Ablamowicz2011}.
 Since $\m{A}_{\bar 1,\bar
2}=\cliffordconjugate{\m{A}}$, the negated MV for $n=2$ algebras
can be expressed through standard involutions.

 (4) The determinant expression for $n=2$ contains only two MVs.
From 8-periodicity table~\cite{Lounesto97} it follows that the
algebras \cl{2}{0}, \cl{1}{1} and \cl{0}{2} are isomorphic to the
algebra $\mathbb{R}(2)$ of real $2\times2$ matrices  or to the
algebra $\mathbb{H}$ of quaternions. The determinant of these
matrices is a quadratic polynomial in the coefficients $a_i$ of
the MVs; therefore, this polynomial can be constructed by
multiplying just two MVs. Below we shall see that in all cases the
total polynomial degree (where coefficients $a_i$ play the role of
variables) of matrix determinant always matches the number of MVs
in the determinant product. The determinant of matrix
representation with quaternionic elements can be calculated using
isomorphism $\mathbb{H}\cong \mathbb{C}(2)$, i.~e. we first
replace quaternions by $2\times2$ block matrices and then
calculate the determinant (the real scalar) of the resulting
matrix. This practical procedure allows us to avoid considering
numerous definitions of the determinant of $\mathbb{H}(n)$ due to
element non-commutativity.

{\bf Three-dimensional quadratic space.}  Our goal is to eliminate
as many grades as possible by forming suitable self-products until
finally the grade-0 element (determinant norm) is left.  As
another example, let us calculate the inverse of $Cl_{2,1}$
algebra. It is easy to check that geometric product of
$\m{A}=a_{0}+a_{1} \e{1}+a_{2} \e{2}+a_{3} \e{3}+a_{4}
\e{12}+a_{5} \e{13}+a_{6} \e{23}+a_{7} \e{123}$ with $\m{A}_{\bar
1,\bar 2}$ lacks grades $1$ and $2$. The result is a new
multivector $\m{B}=\m{A} \m{A}_{\bar 1,\bar 2}= b_{0}+b_{7}
\e{123}$ which  consists of scalar and grade-$3$ element with
coefficients $b_0=a_{0}^2 - a_{1}^2 - a_{2}^2 + a_{3}^2 + a_{4}^2
- a_{5}^2 - a_{6}^2 + a_{7}^2$ and $b_7=-2 a_{3} a_{4} + 2 a_{2}
a_{5} - 2 a_{1} a_{6} + 2 a_{0} a_{7}$. Repeating grade negation
procedure one finds that grade-$3$ part can be removed too, and we
obtain the determinant norm $s_4=\m{B} \m{B}_{\bar 3}=\m{A}
\m{A}_{\bar 1,\bar 2}(\m{A} \m{A}_{\bar 1,\bar 2})_{\bar 3}=b_0^2
- b_7^2$. Reversion of $s_4$ then immediately yields an
alternative form of the norm $s_4^\prime=(\reverse{\m{A}}_{\bar
1,\bar 2}\reverse{\m{A}})_{\bar 3} \reverse{\m{A}}_{\bar 1,\bar
2}\reverse{\m{A}}$. Of course, for different 3D algebras the same
formula will give distinct real expressions, which differ in signs
at individual coefficients~$a_i$. The important thing is that
despite the fact that the scalar $s_4^\prime$ was calculated in
$Cl_{2,1}$ algebra, exactly the same sequence of products and
grade negations will produce the scalar (generally different) in
all other algebras with $n=3$. Also note, that the determinant
norm in this case is determined by two terms, $b_0^2$ and $b_7^2$,
the difference of which should not be equal to zero for an
invertible multivector $\m{A}$ to exist. The condition
$b_0^2-b_7^2\ne 0$ exactly matches the MV invertibility condition,
which according to 8-periodicity table may be obtained by
calculating the determinant of a matrix representation of MV.

{\bf Four-dimensional quadratic space.} In $n=4$ case, in trying
to eliminate as much grades as possible  we can proceed in two
alternative ways. Firstly, we can negate simultaneously the grades
$1$ and $2$ and then in the next step the grades $3$ and $4$.
Alternatively,  we can eliminate $2$ and $3$, and then $1$ and $4$
grades. Both choices are valid. However, if we choose  $1$ and
$3$, and then $2$ and $4$ grade combinations, neither one will do
the job. Thus, we find the following formulas for determinant norm
$s_4= \m{A}\m{A}_{\bar{1},\bar{2}}(\m{A}
\m{A}_{\bar{1},\bar{2}})_{\bar{3},\bar{4}}= \m{A}
\m{A}_{\bar{2},\bar{3}}
(\m{A}\m{A}_{\bar{2},\bar{3}})_{\bar{1},\bar{4}}$ and $s_4^\prime=
(\reverse{\m{A}}_{\bar{1},\bar{2}}\reverse{\m{A}})_{\bar{3},\bar{4}}\,\reverse{\m{A}}_{\bar{1},\bar{2}}\reverse{\m{A}}=(\reverse{\m{A}}_{\bar{2},\bar{3}}\reverse{\m{A}})_{\bar{1},\bar{4}}\reverse{\m{A}}_{\bar{2},\bar{3}}
\reverse{\m{A}}$. It is easy to check that both expressions indeed
give determinant norms.  The occurrence of symbol $\m{A}$ as often
as four times in the products $s_4$ and  $s_4^{\prime}$ is again
what we expect from the  matrix representations for $n=4$. The
total degree of the determinant, considered as a polynomial
function of the  coefficients of MV, is $4$ for all algebras in
the case $n=4$. Despite different forms, the expanded explicit
expressions for $s_m$ and $s_m^\prime$ were found to be equal,
$s_m=s_m^\prime$, as it should be~\cite{Hitzer2016}. Our
computations show that formulas $s_m$ and $s_m^\prime$ are the
only possible equivalent ways to get determinant norm in $n=4$
case using geometric product and negation operations.

{\bf Five-dimensional quadratic space.} This is the largest
dimension, $n=5$, when consecutive grade elimination works by
factorizing the determinant norm into product of the initial MV
and negated ones. The grade elimination sequence is
similar~\cite{Hitzer2016}: (1) Eliminate simultaneously grades $2$
and~$3$; (2) Then eliminate grades $1$ and~$4$; (3) Finally
eliminate grade~$5$. Apart from the last step this sequence is
exactly the same as the second alternative of $n=4$ case. The
final result is
$s_8^\prime=(\m{F}\reverse{\m{A}})_{\bar{5}}\m{F}\reverse{\m{A}}$
with
$\m{F}=(\reverse{\m{A}}_{\bar{2},\bar{3}}\reverse{\m{A}})_{\bar{1},\bar{4}}
\reverse{\m{A}}_{\bar{2},\bar{3}}$ and $s_8=\m{A}\m{D}\,
(\m{A}\m{D})_{\bar{5}}$ with $\m{D}=\m{A}_{\bar{2},\bar{3}}
(\m{A}\m{A}_{\bar{2},\bar{3}})_{\bar{1},\bar{4}}$.

The total degree of determinant polynomial in this  case is $8$,
which again exactly matches the number of MVs in the determinant
norm product. With our program~\cite{AcusDargys2017} we have found
that  $s_8$ determinant norm can be written in 52~ different ways
as presented in Table~\ref{dim5F}.

\begin{table}
{\small
\[\begin{array}{llll}
  N\downarrow&  \textrm{Abbreviation}  \rightarrow&
H=\m{A}\reverse{\m{A}}=\m{A}\m{A}_{\bar{2},\bar{3}}&H^\prime=\m{A}\cliffordconjugate{\m{A}}=\m{A}\m{A}_{\bar{1},\bar{2},\bar{5}} \\[5pt]\hline\\[-5pt]
 13&
    H (H (H_{\bar{1},\bar{4}} H)_{\bar{5}})_{\bar{1},\bar{4}}&
    H (H (H H_{\bar{1},\bar{4}})_{\bar{5}})_{\bar{1},\bar{4}}&
    H (H (HH_{\bar{1},\bar{5}})_{\bar{4}})_{\bar{1},\bar{5}}\\
   &
 H (H(HH_{\bar{4},\bar{5}})_{\bar{1}})_{\bar{4},\bar{5}}&
    H H_{\bar{1},\bar{4}}(H_{\bar{1},\bar{4}}H)_{\bar{5}}&
    \underline{H H_{\bar{1},\bar{4}}(HH_{\bar{1},\bar{4}})_{\bar{5}}}\\
   &
 H H_{\bar{1},\bar{5}}(H_{\bar{1},\bar{5}}H)_{\bar{4}}&
 H H_{\bar{4},\bar{5}}(H_{\bar{4},\bar{5}}H)_{\bar{1}}& \\[5pt]
 14&
 H (H_{\bar{1},\bar{4}}HH_{\bar{1},\bar{5}})_{\bar{4},\bar{5}}&
 H (H_{\bar{1},\bar{4}}HH_{\bar{4},\bar{5}})_{\bar{1},\bar{5}}&
 H (H_{\bar{1},\bar{5}}H_{\bar{1},\bar{4}}H)_{\bar{4},\bar{5}}\\
   &
 H (H_{\bar{1},\bar{5}}H_{\bar{4},\bar{5}}H))_{\bar{1},\bar{4}}&
 H (H_{\bar{1},\bar{5}}HH_{\bar{1},\bar{4}})_{\bar{4},\bar{5}}&
 H (H_{\bar{4},\bar{5}}H_{\bar{1},\bar{4}}H)_{\bar{1},\bar{5}}\\
   &
 H (H_{\bar{4},\bar{5}}H_{\bar{1},\bar{5}}H)_{\bar{1},\bar{4}}&
 H (H_{\bar{4},\bar{5}}HH_{\bar{1},\bar{4}})_{\bar{1},\bar{5}}&
 H (H(H_{\bar{1},\bar{5}}H)_{\bar{3},\bar{4}})_{\bar{1},\bar{5}}\\
   &
 H (H(H_{\bar{4},\bar{5}}H)_{\bar{1},\bar{3}})_{\bar{4},\bar{5}}&
 H (HH_{\bar{1},\bar{4}}H_{\bar{1},\bar{5}})_{\bar{4},\bar{5}}&
 H (HH_{\bar{1},\bar{4}}H_{\bar{4},\bar{5}})_{\bar{1},\bar{5}}\\
   &
 H (HH_{\bar{1},\bar{5}}H_{\bar{4},\bar{5}})_{\bar{1},\bar{4}}&
 H (HH_{\bar{4},\bar{5}}H_{\bar{1},\bar{5}})_{\bar{1},\bar{4}}&
 H H_{\bar{1},\bar{4}} H_{\bar{1},\bar{5}} H_{\bar{4},\bar{5}}\\
   &
 H H_{\bar{1},\bar{4}}H_{\bar{4},\bar{5}}H_{\bar{1},\bar{5}}&
 \underline{H H_{\bar{1},\bar{5}}(HH_{\bar{1},\bar{5}})_{\bar{3},\bar{4}}}&
 H H_{\bar{1},\bar{5}}H_{\bar{4},\bar{5}}H_{\bar{1},\bar{4}}\\
    &
 \underline{H H_{\bar{4},\bar{5}}(HH_{\bar{4},\bar{5}})_{\bar{1},\bar{3}}}&
 H H_{\bar{4},\bar{5}}H_{\bar{1},\bar{5}}H_{\bar{1},\bar{4}} & \\[5pt]
15&
 H (H_{\bar{1},\bar{4}}(H_{\bar{1},\bar{5}}H)_{\bar{3}})_{\bar{4},\bar{5}}&
 H (H_{\bar{1},\bar{4}}(H_{\bar{4},\bar{5}}H)_{\bar{3}})_{\bar{1},\bar{5}}&
 H (H_{\bar{1},\bar{5}}(HH_{\bar{4},\bar{5}})_{\bar{3}})_{\bar{1},\bar{4}}\\
   &
 H (H_{\bar{4},\bar{5}}(HH_{\bar{1},\bar{5}})_{\bar{3}})_{\bar{1},\bar{4}}&
 H (H(H_{\bar{1},\bar{5}}H_{\bar{1},\bar{4}})_{\bar{3}})_{\bar{4},\bar{5}}&
 H (H(H_{\bar{4},\bar{5}}H_{\bar{1},\bar{4}})_{\bar{3}})_{\bar{1},\bar{5}}\\
   &
 H H_{\bar{1},\bar{5}}(H_{\bar{1},\bar{4}}H_{\bar{4},\bar{5}})_{\bar{3}}&
 H H_{\bar{4},\bar{5}}(H_{\bar{1},\bar{4}}H_{\bar{1},\bar{5}})_{\bar{3}} & \\[5pt]
17&
 H (H_{\bar{1},\bar{4}}(H_{\bar{1},\bar{4}}H_{\bar{1},\bar{5}})_{\bar{1}})_{\bar{1},\bar{5}}&
 H (H_{\bar{1},\bar{4}}(H_{\bar{1},\bar{4}}H_{\bar{4},\bar{5}})_{\bar{4}})_{\bar{4},\bar{5}}&
 H (H_{\bar{1},\bar{5}}(H_{\bar{1},\bar{5}}H_{\bar{1},\bar{4}})_{\bar{1}})_{\bar{1},\bar{4}}\\
  &
 H (H_{\bar{1},\bar{5}}(H_{\bar{1},\bar{5}}H_{\bar{4},\bar{5}})_{\bar{5}})_{\bar{4},\bar{5}}&
 H (H_{\bar{1},\bar{5}}(H_{\bar{4},\bar{5}}H_{\bar{1},\bar{5}})_{\bar{5}})_{\bar{4},\bar{5}}&
 H (H_{\bar{4},\bar{5}}(H_{\bar{1},\bar{5}}H_{\bar{4},\bar{5}})_{\bar{5}})_{\bar{1},\bar{5}}\\
  &
 H (H_{\bar{4},\bar{5}}(H_{\bar{4},\bar{5}}H_{\bar{1},\bar{4}})_{\bar{4}})_{\bar{1},\bar{4}}&
 H (H_{\bar{4},\bar{5}}(H_{\bar{4},\bar{5}}H_{\bar{1},\bar{5}})_{\bar{5}})_{\bar{1},\bar{5}} & \\[5pt]
18&
 H (H_{\bar{1},\bar{4}}(H_{\bar{1},\bar{5}}H_{\bar{1},\bar{4}})_{\bar{1},\bar{3}})_{\bar{1},\bar{5}}&
 H (H_{\bar{1},\bar{4}}(H_{\bar{4},\bar{5}}H_{\bar{1},\bar{4}})_{\bar{3},\bar{4}})_{\bar{4},\bar{5}}&
 H (H_{\bar{1},\bar{5}}(H_{\bar{1},\bar{4}}H_{\bar{1},\bar{5}})_{\bar{1},\bar{3}})_{\bar{1},\bar{4}}\\
  &
H (H_{\bar{4},\bar{5}}(H_{\bar{1},\bar{4}}H_{\bar{4},\bar{5}})_{\bar{3},\bar{4}})_{\bar{1},\bar{4}} & & \\[5pt]
15&
 H^\prime (H^\prime (H^\prime H^\prime_{\bar{3}})_{\bar{4}})_{\bar{3}}&
 H^\prime H^\prime_{\bar{3}}(H^\prime_{\bar{3}} H^\prime)_{\bar{4}} & \\[5pt]
16&
 H^\prime (H^\prime (H^\prime_{\bar{3}} H^\prime)_{\bar{1},\bar{4}})_{\bar{3}}&
 H^\prime (H^\prime_{\bar{3}} (H^\prime H^\prime_{\bar{3}})_{\bar{1},\bar{4}} & \\
\end{array}
\]
\caption{\label{dim5F} Alternative formulas for MV determinant
  norm for GAs of vector space dimension $n=5$ listed by increasing number of
negations. For example, the number of negations  $N$ in the first
line is determined by four $H=\m{A}\m{A}_{\bar{2},\bar{3}}$, each
including $2$ negations, $(\bar{2},\bar{3})$, and five explicit
negations $(\bar{1},\bar{4})+\bar{5}+(\bar{1},\bar{4})$ in the
formula, resulting in $2*4+5=13$ total negations. Computationally
preferred forms that contain the largest number of repeating
pieces are underlined.}
  }
\end{table}


\vspace{2mm}
 \textit{Example~1}. Let's take
$\m{A}=3+\e{2}+\e{5}-\e{12}-\e{15}+3 \e{125}$ in $Cl_{4,1}$. It is
easy to check that $\m{A}\m{A}_{\bar{2},\bar{3}}=0$. In
Table~\ref{dim5F}, the formulas with
$H=\m{A}\m{A}_{\bar{2},\bar{3}}$ immediately allow to conclude
that the determinant norm of this MV is zero and therefore  the
inverse of $\m{A}$ does not exist. Now let's try to find the
determinant of $\m{A}$ with formula that does not contain
$\m{A}\m{A}_{\bar{2},\bar{3}}=0$, for example with
$H^\prime=\m{A}\m{A}_{\bar{1},\bar{2},\bar{5}}$ (see the first
line in Table~\ref{dim5F}), from which the final result is not so
obvious. First, we calculate $H^\prime= 18+18 \e{125}$, which is
not zero. However, computing the next step $H^\prime
(H^\prime)_{\bar{3}}$ and $ (H^\prime)_{\bar{3}} H^\prime$ in the
last two lines  in Table~\ref{dim5F}  we get zero again.

\vspace{2mm}
 \textit{Example~2}. Given $\m{A}=1+2 \e{1}+3 \e{23}+4
\e{2345}$ in $Cl_{5,0}$ let us find $\m{A}^{-1}$ using a couple of
alternative formulas. First, we shall use new computationally
efficient formula $D_{1}=H H_{\bar{1},\bar{5}} (H
H_{\bar{1},\bar{5}})_{\bar{3},\bar{4}}$ (underlined in
Table~\ref{dim5F}). Computation of  $H$ yields
$H=\m{A}\m{A}_{\bar{2},\bar{3}}=30+4 \e{1}+8 \e{2345}+16
\e{12345}$. Then $H H_{\bar{1},\bar{5}}=692+352 \e{2345}$. And
lastly, $D_1=354960$. Then the inverse is
$\m{A}^{-1}=\frac{\m{A}_{\bar{2},\bar{3}}H_{\bar{1},\bar{5}} (H
H_{\bar{1},\bar{5}})_{\bar{3},\bar{4}}}{D_1}=
\frac{1}{354960}(3576+96 \e{1}-53832 \e{23}-15072 \e{45}-8592
\e{123}-28992 \e{145}+47424 \e{2345}-8256 \e{12345})$.

Now let's compute the determinant norm of $\m{A}$ using $D_2=H
H_{\bar{4},\bar{5}} H_{\bar{1},\bar{5}} H_{\bar{1},\bar{4}}$,
which computationally is less efficient, because it contains
smaller number of repeating parts. Computation of $H
H_{\bar{4},\bar{5}}$ yields $596-16 \e{1}$. Then, $H
H_{\bar{4},\bar{5}} H_{\bar{1},\bar{5}}$ gives $17944-2864
\e{1}+5024 \e{2345}-9664 \e{12345}$. Finally, $D_2=354960$. Of
course, these formulas give the same explicit expressions for
inverse MV in symbolic form as well.

\vspace{2mm}

Can the above described determinant computation  procedure be
extended beyond $n=5$? The short answer is `yes' if, as shown
below, we allow linear combinations of grade-negated self-products
with properly chosen numerical coefficients. Before describing
this case let us derive some useful formulas for even subalgebras
when $n\le6$. The even subalgebras are directly related with the
spinor groups that are very important in the quantum
mechanics~\cite{Vaz2016}.

\section{Inverse of even multivectors\label{sec:3}}

\begin{table}
\[\begin{array}{ll}
Cl_{p,q}&\qquad  \bigl(\m{A}^{+}\bigr)^{-1} \\[5pt]\hline
p+q=2&\qquad \frac{B \bigl(\m{A}^{+} B\bigr)} {\bigl(\m{A}^{+}
B\bigr)^2}, \qquad B =(v)_{\bar{1}}=-v
\\[4pt]
  p+q=3 &\qquad \frac{C \bigl(\m{A}^{+} C\bigr)}
{\bigl(\m{A}^{+} C\bigr)^2}, \qquad C = (\m{A}^{+}
v)_{\bar{1}}
\\[4pt]
  p+q=4&\qquad \frac{C \bigl(\m{A}^{+} C\bigr)}
{\bigl(\m{A}^{+} C\bigr)^2}, \qquad  C = (\m{A}^{+}
v)_{\bar{1}}
\\[4pt]
 p+q=5&\qquad \frac{D \bigl(\m{A}^{+} D\bigr)}
{\bigl(\m{A}^{+} D\bigr)^2}, \qquad  D = (\m{A}^{+}
v)_{\bar{3}} \bigl(\m{A}^{+} (\m{A}^{+}
v)_{\bar{3}}\bigr)_{\bar{1}}
\\[4pt]
p+q=6&\qquad \frac{D \bigl(\m{A}^{+} D\bigr)_{\bar{6}}}
{\bigl(\m{A}^{+} D\bigr)\bigl(\m{A}^{+} D\bigr)_{\bar{6}}}, \quad
D = (\m{A}^{+} v)_{\bar{3}} \bigl(\m{A}^{+} (\m{A}^{+}
v)_{\bar{3}}\bigr)_{\bar{1}}
\end{array}
\]
\caption{\label{inveven} Explicit formulas of the inverse of even
MVs in a coordinate-free form for Clifford algebras of dimension
$n=p+q\le 6$. The quantities $\bigl(\m{A}^{+} B\bigr)^2/v^2$,
$\bigl(\m{A}^{+} C\bigr)^2/v^2$, $\bigl(\m{A}^{+} D\bigr)^2/v^4$
and ${\bigl(\m{A}^{+} D\bigr)\bigl(\m{A}^{+}
D\bigr)_{\bar{6}}}/v^4$  are the determinant norms of respective
MVs. Here,  nonisotropic and unnormalized  vector $v$ also can be
replaced by one of orthonormal base vector $\e{i}$ for
computational efficiency.}
\end{table}

When  MV consists of even grade elements only, i.~e.
$\m{A}^{+}=\langle\m{A}\rangle_{0}+\langle\m{A}\rangle_{2}+\langle\m{A}\rangle_{4}+\cdots$,
simpler formulas for inverse MVs can be constructed as shown in
Table~\ref{inveven}. The nonisotropic unnormalized vector $v$ in these formulas play the
role of dummy variable.

It is interesting to observe that in contrast to general case
considered in the next section there exists a single self-negated
product for inverse of even MV (see Table~\ref{inveven}) that is
not a linear combination. This property is to be expected (compare
with  a general MV form for $n=5$ in Table~\ref{invall}), because
there exists the well-known isomorphism between the even
subalgebra of one vector space dimension larger algebra  and the
full lower dimensional Clifford algebra,
\begin{equation}\label{iso}
\cl{p}{q+1}^+\cong\cl{p}{q},\quad\cl{p+1}{q}^+\cong\cl{q}{p}\,.
\end{equation}
Because we can write inverse MV as a single self-negated product
for dimension $n=5$ (see Table~\ref{invall}), it is not surprising
that according to isomorphisms~\eqref{iso} we can do this for even
subalgebra of dimension $n=6$. Unfortunately, as we shall see in
section~\ref{sec:4} this property does not extend to the full six
dimensional Clifford algebra.

\section{Inverse of general MV in 6-dimensional quadratic space\label{sec:4}}
\subsection{Insufficiency of single self-negated product in $n=6$ case\label{sec:4a}}

Let us take $Cl_{6,0}$ algebra. Using \textit{Mathematica} GA
package~\cite{AcusDargys2017}, after some experimentation with a
pair of self-negated product formed from a general MV it is not
difficult to ascertain that we can eliminate simultaneously either
grades $1$, $2$, $5$ and $6$ (then $0$, $3$ and $4$ grades
survive) or, alternatively, the grades $2$, $3$ and $6$ (then the
grades $0$, $1$, $4$ and $5$ survive). In both cases the
grade $4$ remains and, therefore, cannot be eliminated by the
method of simple self-negated product used till now. This
conclusion strictly follows from an attempt to simultaneously
nullify all coefficients of grade-$4$ base elements using all
$2^6=64$ possible combinations of grade negations in the two term
product. The grade-$4$ therefore is distinct from the other grades
and deserves special examination. From experiments by computer it
turns out that there exists a subalgebra with base formed by
elements $\{1, \e{1256}, \e{1346},\e{2345}\}$ such that any self
product of multivector $\m{B}=a_{0}+a_{47} \e{1256}+a_{49}
\e{1346}+a_{52} \e{2345}$  by grade-negated $\m{B}$ will yield new
MV having at least one grade-$4$ base element present. This is the
reason why such a single self-product fails in eliminating
grade-$4$ part and, as we shall see, one has to use a combination
of at least a pair of self-products.

\subsection{Linear combination of self-products\label{sec:4b}}
Before considering general case, we shall note that the
determinant of a matrix representation (computed using symbolic
coefficients) of restricted MV, which consists of just scalar and
general grade-4 element, can be written as a square of some
polynomial of MV coefficients $s_4$, i.~e. as $\det
\bigl(\langle\m{A}\rangle_{0}+\langle\m{A}\rangle_{4}\bigr)=s_8=(s_4)^2$.
Thus, we assume that one can always extract the square root
from~$s_8$. From this follows that in search of inverse of
$\langle\m{A}\rangle_{0}+\langle\m{A}\rangle_{4}$,
 as a first step one can try to test only
linear combination of product of four negated MVs instead of
eight as required in general case.  Due to above arguments we
assume that square root of determinant norm may be written as a
linear combination of the following form
\begin{equation}\begin{split}
 &s_{4+4}=s_{4f}+s_{4g}=\\
 &b_1 B f_5\Bigl(f_4(B) f_3\bigl(f_2(B) f_1(B)\bigr)\Bigr)
 +b_2 B g_5\Bigl(g_4(B) g_3\bigl(g_2(B)
  g_1(B)\bigr)\Bigr)\label{s1},
\end{split}\end{equation}
where  each of $f_j$ and $g_j$ is either the identity mapping or
the grade-$4$ negation, and $b_1, b_2$ are unknown scalar
coefficients of the linear combination.

Once a pattern of linear combination of square root of the
determinant norm was fixed we can calculate explicit symbolic form of
matrix representation of the above mentioned MV
$\m{B}=a_{0}+a_{47} \e{1256}+a_{49} \e{1346}+a_{52} \e{2345}$,
then compute the matrix determinant, take square root and compare
it with the GA expression~\eqref{s1} after the same GA multivector
$\m{B}$ was inserted.

The negation functions $f_j$ and $g_j$ that control signs of
grades can be modelled as a multiplication by unknown coefficient
$p_{4jk}$, where the index $j$ denotes the involution number
$f_j,g_j$ in the self-product and $k$ is the term number in linear
combination (when $n=6$, $k=1$ for $f$ and $k=2$ for $g$). Later,
when we shall deal with negations of other grades the first index
$i$ in $p_{ijk}$ will indicate possibly negated grade-$i$. For the
moment the index $i$ is fixed to $4$, which corresponds to current
nontrivial grade of $\m{B}$ (we will ignore negations of scalar,
because it is equivalent to negation of all other remaining MV
grades). The coefficients $p_{ijk}$ acquire  values $\pm 1$ only,
where $-1$ means that involution which changes sign of grade $i$
is to be applied, while  $+1$ means the identity map.

In the considered $n=6$ case, comparison with the square root of
the determinant of a matrix representation of $\m{B}$ yields the
system of four equations for each of base element (including the
scalar) of $\m{B}$. The system is too long to be fully presented
here, therefore  a small characteristic part of it is written down
in a truncated form below,
\begin{equation}
\left\{
\begin{array}{l}
a_{0}^4 - 2 a_{0}^2 a_{47}^2 + b_2 a_{0}^2 a_{47}^2 p_{4 1 2} p_{4 2 2}+ <\textrm{72 monomials}>= 0\\
b_1 a_{0}^3 a_{52}+2b_2 a_0 a_{47}^2 a_{52} p_{412} p_{422} p_{4 3 2} p_{4 4 2} p_{4 52}
+ <\textrm{72 monomials}>= 0\\
<\textrm{74 monomials}>=0\\
<\textrm{74 monomials}>=0
\end{array}\right.\label{explicitEq}
\end{equation}
We see that even for a simple $\m{B}$ which contains only 4~base
elements (the scalar, $\e{1256}$, $\e{1346}$ and $\e{2345}$) the
system is highly nonlinear in $p_{4ij}$. We can, however, try to
substitute concrete values for $p_{4ij}$ one by one to get much
simpler systems that  contain only variables $b_1, b_2$ and $a_i$.
Then we can try to solve each of simple systems separately with
respect to $b_1, b_2$ for arbitrary coefficients $a_i$. Only few
of them have solutions. In fact, we have solved the systems with
{\it Mathematica} command \textbf{SolveAlways[~]}. After testing
all $2^{10}=1024$ possible values of $p_{4ij}$ we have found two
sets of solutions, $\{b_1=-2/3,\; b_2=-1/3,\; p_{411}=-1,\;
p_{412}=1,\; p_{421}=-1,\; p_{422}=1,\; p_{431}=-1,\;
p_{432}=-1,\; p_{441}=-1,\; p_{442}=1,\; p_{451}=-1,\;
p_{452}=1\}$ and $\{b_1=-1/3,\; b_2=-2/3,\; p_{411}=1,\;
p_{412}=-1,\; p_{421}=1,\; p_{422}=-1,\; p_{431}=-1,\;
p_{432}=-1,\; p_{441}=1,\; p_{442}=-1,\; p_{451}=1,\;
p_{452}=-1\}$. The obtained solution is unique up to the
permutation of two terms. The common sign of coefficients $b_1$
and $b_2$ in general is not fixed as yet, because  square root of
determinant norm was calculated at this stage.

It is easy to check that the above solution (though computed from
highly simplified MV, namely, the scalar plus three grade-$4$ base
elements) works flawlessly for a more general MV (scalar plus any
number of grade-$4$ elements) $\m{C}= a_0+a_{42} \e{1234}+a_{43}
\e{1235}+a_{44} \e{1236}+a_{45} \e{1245}+a_{46} \e{1246}+a_{47}
\e{1256}+a_{48} \e{1345}+a_{49} \e{1346}+a_{50} \e{1356}+a_{51}
\e{1456}+a_{52} \e{2345}+a_{53} \e{2346}+a_{54} \e{2356}+a_{55}
\e{2456}+a_{56} \e{3456} $. This is what one expects, because
grade involution done on the same  grade elements acts in exactly
identical way. Starting with a simple MV and then augmenting it
till the number of solutions cannot be further decreased allows
one to keep the whole calculation size manageable. This
considerably speeds up search procedure for involutions and
balances calculation complexity against the speed.

Once  $f_j, g_j$ and coefficients  $b_k$ in
equations~\eqref{explicitEq} and \eqref{s1} have been
determined, we can renew the search for other negation
involutions in exactly the same way. Our next task is to double
the number of multivectors in the products~\eqref{s1}, because we
know that determinant of general MV requires 8~multipliers in
order to match the total degree polynomial of determinant of a MV
matrix representation.

It was already mentioned that in the product
$\m{A}\m{A}_{i,j,...}$ one can simultaneously eliminate either
grades $1$, $2$, $5$ and $6$ (then grades $0$, $3$ and $4$
remain), or, alternatively, the grades $2$, $3$ and $6$ (then
grades $0$, $1$, $4$ and $5$ survive). All in all there are
$\frac{6!}{2!4!}+\frac{6!}{3!3!} +1=36$ elements of grades $2$,
$3$ and $6$ which can be eliminated simultaneously. This is more
than $28$  elements of grades $1$, $2$, $5$ and $6$.  Therefore,
if we replace $\m{B}$ in equation~\eqref{s1} by self-negated
product $\m{A}\m{A}_{{\bar 2},{\bar 3},{\bar 6}}$ we are left to
deal with self-negated product of four multivectors of grades $0$,
$1$, $4$, $5$ only, from which we can ignore  the grades $0$ and
$4$ for which $f_j$ and $g_j$ already have been established. Thus,
repeating the same procedure we can find  a number of valid
solutions for coefficients $p_{1jk},p_{2jk}$ and $p_{5jk}$. In
order to speed up the derivation we had used the multivector $
a_{0}+a_{1} \e{1}+a_{22} \e{123}+a_{47} \e{1256}$ first. The
obtained result then was explicitly verified (i.~e. proved by
explicit expansion in orthogonal basis) using the most general
$Cl_{6,0}$ multivector with symbolic coefficients. Finally we
found $320$ valid forms of determinant norm. After removing all
superfluous involutions a number of possible determinant forms was
reduced to 16+4=20. All these forms are presented in
Table~\ref{dim6F}.

After  the determinant norm has been found  the inverse multivector can
be immediately written using equation~\eqref{inverse}. The results
for inverse MVs for algebras $n\le 6$ are summarized in
Table~\ref{invall}.
\begin{table}
\[\begin{array}{ll}
\cl{p}{q}&\qquad  \m{A}^{-1} \\[5pt]\hline\\[-5pt]
p+q=0&\qquad\frac{\m{B}}{\m{A}\m{B}},\qquad\qquad\qquad\qquad\m{B}=1\\[4pt]
 p+q=1&\qquad
\frac{\m{B}(\m{A}\m{B})_{\bar{1}}}{\m{A}\m{B}\,(\m{A}\m{B})_{\bar{1}}}
= \frac{\gradeinverse{\m{A}}}{\m{A} \gradeinverse{\m{A}}},\quad\qquad\ \m{B}=1\\[4pt]
  p+q=2&\qquad \frac{\m{C}}{\m{A}\m{C}}=\frac{\cliffordconjugate{\m{A}}}{\m{A} \cliffordconjugate{\m{A}}}
,\qquad\qquad\quad\ \m{C}=\m{A}_{\bar{1},\bar{2}}\\[4pt]
  p+q=3 &\qquad
 \frac{\m{C}(\m{A}\m{C})_{\bar{3}}}{\m{A}\m{C}\,(\m{A}\m{C})_{\bar{3}}}
=\frac{\cliffordconjugate{\m{A}}\gradeinverse{\m{A}}\reverse{\m{A}}}{\m{A}
\cliffordconjugate{\m{A}}\gradeinverse{\m{A}}\reverse{\m{A}}}
,\qquad\ \m{C}=\m{A}_{\bar{1},\bar{2}}\\[4pt]
 p+q=4&\qquad\frac{\m{D}}{\m{A}\m{D}},\qquad\qquad\qquad\qquad
\m{D}=\m{A}_{\bar{2},\bar{3}}
(\m{A}\m{A}_{\bar{2},\bar{3}})_{\bar{1},\bar{4}}\quad \\
   &\qquad\qquad\qquad\qquad\qquad\textrm{or}\quad \m{D}=\m{A}_{\bar{1},\bar{2}}(\m{A}
\m{A}_{\bar{1},\bar{2}})_{\bar{3},\bar{4}}  \\[4pt]
 p+q=5&\qquad\frac{\m{D}(\m{A}\m{D})_{\bar{5}}}{\m{A}\m{D}\,
(\m{A}\m{D})_{\bar{5}}},\qquad\qquad\qquad
\m{D}=\m{A}_{\bar{2},\bar{3}}
(\m{A}\m{A}_{\bar{2},\bar{3}})_{\bar{1},\bar{4}}\\[4pt]
p+q=6&\qquad\frac{\m{G}}{\m{A}\m{G}},\qquad\m{G}=\frac{1}{3} \m{A}_{\bar{2},\bar{3},\bar{6}} \Bigl(\m{H} (\m{H} \m{H})_{\bar{1},\bar{4},\bar{5}} + 2 \bigl(\m{H}_{\bar{4}} (\m{H}_{\bar{4}} \m{H}_{\bar{4}})_{\bar{1},\bar{4},\bar{5}}\bigr)_{\bar{4}}\Bigr)\\
     &\quad\textrm{or}\quad \m{G}=\frac{1}{3} \m{A}_{\bar{2},\bar{3},\bar{6}} \Bigl(\bigl(\m{H} (\m{H} \m{H}_{\bar{1},\bar{5}})_{\bar{4}}\bigr)_{\bar{1},\bar{5}} + 2 \bigl(\m{H}_{\bar{4},\bar{5}} (\m{H}_{\bar{4},\bar{5}} \m{H}_{\bar{1},\bar{4}})_{\bar{4}}\bigr)_{\bar{1},\bar{4}}\Bigr)\\
&\qquad \qquad \qquad\textrm{with}\quad \m{H}=\m{A} \m{A}_{\bar{2},\bar{3},\bar{6}}=\m{A}\reverse{\m{A}}
\end{array}
\]
\caption{\label{invall}Summary of formulas for inverse
multivectors  $\m{A}^{-1}$ in  Clifford algebras of dimension
$n\le 6$.}
\end{table}

The denominators, which are real scalars, in these expressions are
the determinant norms of respective MVs. At the same time they
give the condition for existence of the inverse MV. It was found
that determinant norms exactly match expressions for determinants
calculated from matrix representation of respective MVs (in fact,
the whole derivation of the determinant norm formulas relied on
this match). All formulas were proved  by explicit calculation of
symbolic inverses of the most general MV for concrete Clifford
algebras with all possible signatures $(p,q)$. For this task we
wrote the {\it Mathematica} package for symbolic calculations in
Clifford algebras~\cite{AcusDargys2017}, which also can be
found in {\it Mathematica} package data base
\url{http://packagedata.net/}. The specific property of our
GA program is that it can simultaneously work with a number of
Clifford algebras having different signatures $(p,q)$ what was a
great help in computer-assisted verifications.

\begin{table}
\[\begin{array}{ll}
  N \downarrow&  \textrm{Abbreviation} \rightarrow H=A\reverse{A}=AA_{\bar{2},\bar{3},\bar{6}} \\[5pt]\hline\\[-5pt]
38&\frac{1}{3} H (H (H H_{\bar{1},\bar{5}})_{\bar{4}})_{\bar{1},\bar{5}}+\frac{2}{3} H (H_{\bar{1},\bar{4}} (H_{\bar{1},\bar{4}} H_{\bar{4},\bar{5}})_{\bar{4}})_{\bar{4},\bar{5}}\\
&\frac{1}{3} H H_{\bar{1},\bar{5}} (H_{\bar{1},\bar{5}} H)_{\bar{4}}+\frac{2}{3} H (H_{\bar{1},\bar{4}} (H_{\bar{1},\bar{4}} H_{\bar{4},\bar{5}})_{\bar{4}})_{\bar{4},\bar{5}}\\
&\frac{1}{3} H (H (H H_{\bar{1},\bar{5}})_{\bar{4}})_{\bar{1},\bar{5}}+\frac{2}{3} H (H_{\bar{4},\bar{5}} (H_{\bar{4},\bar{5}} H_{\bar{1},\bar{4}})_{\bar{4}})_{\bar{1},\bar{4}}\\
&\frac{1}{3} H H_{\bar{1},\bar{5}} (H_{\bar{1},\bar{5}} H)_{\bar{4}}+\frac{2}{3} H (H_{\bar{4},\bar{5}} (H_{\bar{4},\bar{5}} H_{\bar{1},\bar{4}})_{\bar{4}})_{\bar{1},\bar{4}}\\[5pt]
 & \\
34&\frac{1}{3} H H (H H)_{\bar{1},\bar{4},\bar{5}}+\frac{2}{3} H ((H)_{\bar{4}} ((H)_{\bar{4}} (H)_{\bar{4}})_{\bar{1},\bar{4},\bar{5}})_{\bar{4}}\\
 & \\
36&\frac{1}{3} H H (H H)_{\bar{1},\bar{4},\bar{5}}+\frac{2}{3} H ((H)_{\bar{4}} ((H)_{\bar{1},\bar{4},\bar{5}} (H)_{\bar{1},\bar{4},\bar{5}})_{\bar{4}})_{\bar{4}}\\
&\frac{1}{3} H H (H H)_{\bar{1},\bar{4},\bar{5}}+\frac{2}{3} H ((H)_{\bar{1},\bar{4},\bar{5}} ((H)_{\bar{4}} (H)_{\bar{4}})_{\bar{4}})_{\bar{1},\bar{4},\bar{5}}\\
&\frac{1}{3} H (H_{\bar{1},\bar{5}} (H H)_{\bar{4}})_{\bar{1},\bar{5}}+\frac{2}{3} H ((H)_{\bar{4}} ((H)_{\bar{4}} (H)_{\bar{4}})_{\bar{1},\bar{4},\bar{5}})_{\bar{4}}\\
&\frac{1}{3} H H (H_{\bar{1},\bar{5}} H_{\bar{1},\bar{5}})_{\bar{4}}+\frac{2}{3} H ((H)_{\bar{4}} ((H)_{\bar{4}} (H)_{\bar{4}})_{\bar{1},\bar{4},\bar{5}})_{\bar{4}}\\
 & \\
38&\frac{1}{3} H (H_{\bar{1},\bar{5}} (H H)_{\bar{4}})_{\bar{1},\bar{5}}+\frac{2}{3} H ((H)_{\bar{4}} ((H)_{\bar{1},\bar{4},\bar{5}} (H)_{\bar{1},\bar{4},\bar{5}})_{\bar{4}})_{\bar{4}}\\
&\frac{1}{3} H H (H_{\bar{1},\bar{5}} H_{\bar{1},\bar{5}})_{\bar{4}}+\frac{2}{3} H ((H)_{\bar{4}} ((H)_{\bar{1},\bar{4},\bar{5}} (H)_{\bar{1},\bar{4},\bar{5}})_{\bar{4}})_{\bar{4}}\\
&\frac{1}{3} H (H_{\bar{1},\bar{5}} (H H)_{\bar{4}})_{\bar{1},\bar{5}}+\frac{2}{3} H ((H)_{\bar{1},\bar{4},\bar{5}} ((H)_{\bar{4}} (H)_{\bar{4}})_{\bar{4}})_{\bar{1},\bar{4},\bar{5}}\\
&\frac{1}{3} H H (H_{\bar{1},\bar{5}} H_{\bar{1},\bar{5}})_{\bar{4}}+\frac{2}{3} H ((H)_{\bar{1},\bar{4},\bar{5}} ((H)_{\bar{4}} (H)_{\bar{4}})_{\bar{4}})_{\bar{1},\bar{4},\bar{5}}\\
 & \\
42&\frac{1}{3} H H (H H)_{\bar{1},\bar{4},\bar{5}}+\frac{2}{3} H ((H)_{\bar{1},\bar{4},\bar{5}} ((H)_{\bar{1},\bar{4},\bar{5}} (H)_{\bar{1},\bar{4},\bar{5}})_{\bar{1},\bar{4},\bar{5}})_{\bar{1},\bar{4},\bar{5}}\\
&\frac{1}{3} H (H_{\bar{1},\bar{5}} (H_{\bar{1},\bar{5}} H_{\bar{1},\bar{5}})_{\bar{1},\bar{4},\bar{5}})_{\bar{1},\bar{5}}+\frac{2}{3} H ((H)_{\bar{4}} ((H)_{\bar{4}} (H)_{\bar{4}})_{\bar{1},\bar{4},\bar{5}})_{\bar{4}}\\
 & \\
44&\frac{1}{3} H (H_{\bar{1},\bar{5}} (H H)_{\bar{4}})_{\bar{1},\bar{5}}+\frac{2}{3} H ((H)_{\bar{1},\bar{4},\bar{5}} ((H)_{\bar{1},\bar{4},\bar{5}} (H)_{\bar{1},\bar{4},\bar{5}})_{\bar{1},\bar{4},\bar{5}})_{\bar{1},\bar{4},\bar{5}}\\
&\frac{1}{3} H H (H_{\bar{1},\bar{5}} H_{\bar{1},\bar{5}})_{\bar{4}}+\frac{2}{3} H ((H)_{\bar{1},\bar{4},\bar{5}} ((H)_{\bar{1},\bar{4},\bar{5}} (H)_{\bar{1},\bar{4},\bar{5}})_{\bar{1},\bar{4},\bar{5}})_{\bar{1},\bar{4},\bar{5}}\\
&\frac{1}{3} H (H_{\bar{1},\bar{5}} (H_{\bar{1},\bar{5}} H_{\bar{1},\bar{5}})_{\bar{1},\bar{4},\bar{5}})_{\bar{1},\bar{5}}+\frac{2}{3} H ((H)_{\bar{4}} ((H)_{\bar{1},\bar{4},\bar{5}} (H)_{\bar{1},\bar{4},\bar{5}})_{\bar{4}})_{\bar{4}}\\
&\frac{1}{3} H (H_{\bar{1},\bar{5}} (H_{\bar{1},\bar{5}} H_{\bar{1},\bar{5}})_{\bar{1},\bar{4},\bar{5}})_{\bar{1},\bar{5}}+\frac{2}{3} H ((H)_{\bar{1},\bar{4},\bar{5}} ((H)_{\bar{4}} (H)_{\bar{4}})_{\bar{4}})_{\bar{1},\bar{4},\bar{5}}\\
 & \\
50&\frac{1}{3} H (H_{\bar{1},\bar{5}} (H_{\bar{1},\bar{5}}
H_{\bar{1},\bar{5}})_{\bar{1},\bar{4},\bar{5}})_{\bar{1},\bar{5}}+\frac{2}{3}
H ((H)_{\bar{1},\bar{4},\bar{5}} ((H)_{\bar{1},\bar{4},\bar{5}}
(H)_{\bar{1},\bar{4},\bar{5}})_{\bar{1},\bar{4},\bar{5}})_{\bar{1},\bar{4},\bar{5}}
\end{array}
\]
\caption{\label{dim6F} The formulas for determinant norm of MV
$\m{A}$ for Clifford algebras of vector space dimension $n=6$. $N$
counts the total number of negations in the
 norm (see Table~\ref{dim5F} for explanation).}
\end{table}

\section{Classification and examples\label{sec:5}}

In Table~\ref{dim6F} we have collected all formulas obtained by
the described method that represent the determinant norm at $n=6$
in the pattern~\eqref{s1}. All expressions for determinant norm
naturally split into two types. The first four formulas at the
beginning of Table~\ref{dim6F} ($N=38$) constitute the first type
or class. All four expressions from this class can be obtained by
forming pairs from two sets
\[\begin{split}&\lbrace\frac{1}{3} H (H (H
H_{\bar{1},\bar{5}})_{\bar{4}})_{\bar{1},\bar{5}},
 \ \frac{1}{3}H H_{\bar{1},\bar{5}} (H_{\bar{1},\bar{5}} H)_{\bar{4}}\rbrace,\quad\text{and}\\
&\{\frac{2}{3} H (H_{\bar{1},\bar{4}} (H_{\bar{1},\bar{4}}
H_{\bar{4},\bar{5}})_{\bar{4}})_{\bar{4},\bar{5}},  \frac{2}{3} H
(H_{\bar{4},\bar{5}} (H_{\bar{4},\bar{5}}
H_{\bar{1},\bar{4}})_{\bar{4}})_{\bar{1},\bar{4}}\} \end{split}\]
in any combination. This gives the first four, $2\times 2=4$,
formulas. The remaining 16 formulas, which constitute the second
class can be obtained  similarly by forming all possible pairs
from  sets
 {\small\[\begin{split}
 &\{\frac{1}{3} H H (H H)_{\bar{1},\bar{4},\bar{5}},\
\frac{1}{3} H (H_{\bar{1},\bar{5}} (H
H)_{\bar{4}})_{\bar{1},\bar{5}}, \frac{1}{3} H H
(H_{\bar{1},\bar{5}} H_{\bar{1},\bar{5}})_{\bar{4}},\\
&\frac{1}{3} H (H_{\bar{1},\bar{5}} (H_{\bar{1},\bar{5}}
H_{\bar{1},\bar{5}})_{\bar{1},\bar{4},\bar{5}})_{\bar{1},\bar{5}}\},
\end{split}\]} and
 {\small\[\begin{split}
 &\{\frac{2}{3} H ((H)_{\bar{4}} ((H)_{\bar{4}}
(H)_{\bar{4}})_{\bar{1},\bar{4},\bar{5}})_{\bar{4}},\frac{2}{3} H
((H)_{\bar{1},\bar{4},\bar{5}} ((H)_{\bar{4}}
(H)_{\bar{4}})_{\bar{4}})_{\bar{1},\bar{4},\bar{5}},\\
&\frac{2}{3} H ((H)_{\bar{4}} ((H)_{\bar{1},\bar{4},\bar{5}}
(H)_{\bar{1},\bar{4},\bar{5}})_{\bar{4}})_{\bar{4}}, \frac{2}{3} H
((H)_{\bar{1},\bar{4},\bar{5}} ((H)_{\bar{1},\bar{4},\bar{5}}
(H)_{\bar{1},\bar{4},\bar{5}})_{\bar{1},\bar{4},\bar{5}})_{\bar{1},\bar{4},\bar{5}}\}.
\end{split}\]}
This yields remaining $4\times 4=16$ formulas.

Both classes are well defined, because symbolic expressions for
each of separate terms in a pair always coincide within the class
but differ between different classes. The representatives of both
classes were included in Table~\ref{invall}. In the first class
only grade-3 and grade-4 cancellation can occur
between both terms in a pair. In the second class, the MV in each
pair  generally is made up of grades 1, 4 and 5, which cancel out
in the final result (see \textit{Example}~4). It is also
interesting to note that formulas of Table~\ref{dim6F} can also be
rewritten as a sum of three different terms  with all weight
coefficients being the same and equal to $1/3$ as shown in
Table~\ref{treetermformula}. The inverse MV formula can be
constructed by taking any three expressions either from sets
$S_1$, $S_2$ and $S_3$, or from sets $T_1$, $T_2$ and $T_3$ listed
in Table~\ref{treetermformula}. For example, the formulas
\[\begin{split}
&\frac{1}{3}  H (H_{\bar{4}} (H_{\bar{4}} H_{\bar{4}}
)_{\bar{1},\bar{4},\bar{5}} )_{\bar{4}}+\frac{1}{3}  H
((H_{\bar{4}} H_{\bar{4}} )_{\bar{4}} H_{\bar{1},\bar{4},\bar{5}}
)_{\bar{1},\bar{4},\bar{5}}+\frac{1}{3}  H H (H H
)_{\bar{1},\bar{4},\bar{5}}\quad \mathrm{and}\\
 &\frac{1}{3} H ( H_{\bar{4},\bar{5}}  (H_{\bar{4},\bar{5}}
H_{\bar{1},\bar{4}} )_{\bar{4}} )_{\bar{1},\bar{4}}
+\frac{1}{3} H (( H_{\bar{1},\bar{4}} H_{\bar{4},\bar{5}} )_{\bar{4}} H_{\bar{4},\bar{5}}
   )_{\bar{1},\bar{4}}
+\frac{1}{3} H(H(H H_{\bar{1},\bar{5}} )_{\bar{4}}
)_{\bar{1},\bar{5}}\end{split}\] are pretty valid choices.

Generally the Table~\ref{treetermformula} allows to make $4^3=64$
different triplets from sets $S_i$ and $2^3=8$ triplets from sets
$T_i$. Thus, all in all we can construct $64+8=72$ different
triplet formulas of inverse for $n=6$ algebras without
taking into account possible reversed forms.

\begin{table}
\[\begin{array}{ll}
S_1=&\{\frac{1}{3}  H (H_{\bar{4}} (H_{\bar{4}} H_{\bar{4}}
)_{\bar{1},\bar{4},\bar{5}} )_{\bar{4}},\quad
\frac{1}{3}  H (H_{\bar{4}} (H_{\bar{1},\bar{4},\bar{5}} H_{\bar{1},\bar{4},\bar{5}} )_{\bar{4}} )_{\bar{4}},\\
&\quad\frac{1}{3}  H (H_{\bar{1},\bar{4},\bar{5}} (H_{\bar{4}}
H_{\bar{4}} )_{\bar{4}} )_{\bar{1},\bar{4},\bar{5}},\quad
\frac{1}{3} H (H_{\bar{1},\bar{4},\bar{5}}
(H_{\bar{1},\bar{4},\bar{5}} H_{\bar{1},\bar{4},\bar{5}}
)_{\bar{1},\bar{4},\bar{5}} )_{\bar{1},\bar{4},\bar{5}}\},
\\
S_2=&\{\frac{1}{3}  H ((H_{\bar{4}} H_{\bar{4}} )_{\bar{1},\bar{4},\bar{5}} H_{\bar{4}} )_{\bar{4}},
\quad \frac{1}{3}  H ((H_{\bar{4}} H_{\bar{4}} )_{\bar{4}} H_{\bar{1},\bar{4},\bar{5}} )_{\bar{1},\bar{4},\bar{5}},\\
 &\quad\frac{1}{3}  H ((H_{\bar{1},\bar{4},\bar{5}} H_{\bar{1},\bar{4},\bar{5}} )_{\bar{4}} H_{\bar{4}} )_{\bar{4}},
 \quad\frac{1}{3}  H ((H_{\bar{1},\bar{4},\bar{5}} H_{\bar{1},\bar{4},\bar{5}} )_{\bar{1},\bar{4},\bar{5}} H_{\bar{1},\bar{4},\bar{5}}
 )_{\bar{1},\bar{4},\bar{5}}\},
\\
S_3=&\{\frac{1}{3}  H (H_{\bar{1},\bar{5}} (H_{\bar{1},\bar{5}} H_{\bar{1},\bar{5}} )_{\bar{1},\bar{4},\bar{5}} )_{\bar{1},\bar{5}},
\quad\frac{1}{3}  H (H_{\bar{1},\bar{5}} (H H )_{\bar{4}} )_{\bar{1},\bar{5}},\\
   &\quad \frac{1}{3}  H H (H_{\bar{1},\bar{5}} H_{\bar{1},\bar{5}} )_{\bar{4}}, \frac{1}{3}  H H (H H
   )_{\bar{1},\bar{4},\bar{5}}\}.
\\[5pt]
T_1=&\{\frac{1}{3}  H (H_{\bar{1},\bar{4}} (H_{\bar{1},\bar{4}}
H_{\bar{4},\bar{5}} )_{\bar{4}} )_{\bar{4},\bar{5}},\quad
\frac{1}{3} H ( H_{\bar{4},\bar{5}}  (H_{\bar{4},\bar{5}}
H_{\bar{1},\bar{4}} )_{\bar{4}} )_{\bar{1},\bar{4}}\},
     \\
T_2=&\{
   \frac{1}{3} H (( H_{\bar{1},\bar{4}} H_{\bar{4},\bar{5}} )_{\bar{4}} H_{\bar{4},\bar{5}}
   )_{\bar{1},\bar{4}},\quad
   \frac{1}{3} H ((H_{\bar{4},\bar{5}}  H_{\bar{1},\bar{4}} )_{\bar{4}} H_{\bar{1},\bar{4}} )_{\bar{4},\bar{5}}\},\\
T_3=&\{
  \frac{1}{3} H(H(H H_{\bar{1},\bar{5}} )_{\bar{4}} )_{\bar{1},\bar{5}},\quad
  \frac{1}{3} H H_{\bar{1},\bar{5}} (H_{\bar{1},\bar{5}} H)_{\bar{4}}\}.
 \end{array}
\]
\caption{The sets for construction of different triplets  of
weight $\tfrac{1}{3}$ in inversion formulas for $n=6$. For details
see text.\label{treetermformula}}
\end{table}

\vspace{2mm}
 \textit{Example~3}. In $Cl_{4,2}$ let's take $\m{A}=2+\e{1}+\e{5}-2
\e{15}+3 \e{26}+3 \e{1256}$ and  find the inverse
with formula from Table~\ref{invall},
$\m{A}\m{G}=\frac{1}{3} H H (H
H)_{\bar{1},\bar{4},\bar{5}}+\frac{2}{3} H (H_{\bar{4}}
(H_{\bar{4}} H_{\bar{4}})_{\bar{1},\bar{4},\bar{5}})_{\bar{4}}$,
which has a minimal number of negations. We obtain
$H=\m{A}\m{A}_{\bar{2},\bar{3},\bar{6}}=8 \e{1}+8 \e{5}$. In the
next step, however, we get zero since  $H H = 0$ as
well as $H_{\bar{4}} H_{\bar{4}} = 0$. In fact, the MV in
the considered example was constructed multiplying the MV $1+2
\e{1}+3 \e{126}$ by isotropic vector $(\e{1}+\e{5})$ from the
left.

It is easy to check that the last mentioned MV $\m{A}^{\prime}=1+2
\e{1}+3 \e{126}$ is non-invertible as well. Indeed,
$H=\m{A}^{\prime}\m{A}^{\prime}_{\bar{2},\bar{3},\bar{6}}=-4+4
\e{1}$, then $H H =32-32 \e{1}$, and finally $H H (H
H)_{\bar{1},\bar{4},\bar{5}}=0$. In a similar way $H_{\bar{4}}
(H_{\bar{4}} H_{\bar{4}})_{\bar{1},\bar{4},\bar{5}}=0$. One can
check that all formulas in Table~\ref{dim6F} will yield the same
result, namely, zero. In \cite{Helmstetter2014} one can find
more information on MVs that contain isotropic multipliers.

\vspace{5mm}
 \textit{Example~4}. In $Cl_{1,5}$ let's find inverse
of $\m{A}=2+\e{1}+4 \e{3}+\e{15}+3 \e{126}$ using the  same
generic formula with minimal number of negations.  Computation
steps are:

 {\small 1) $H=\m{A}\m{A}_{\bar{2},\bar{3},\bar{6}}=-3+4
\e{1}+16 \e{3}-2 \e{5}-24 \e{1236}$,

2)$H H = -811-24 \e{1}-96 \e{3}+12 \e{5}+144 \e{1236}-96
\e{12356}$,

3)$\frac{1}{3} H H (H H)_{\bar{1},\bar{4},\bar{5}}=1/3
(678025+27648 \e{5}+2304 \e{1236}+18432 \e{1256}-4608
\e{2356}+3456 \e{12356})$,

4) $(H_{\bar{4}} H_{\bar{4}})_{\bar{1},\bar{4},\bar{5}}=-811+24
\e{1}+96 \e{3}-12 \e{5}+144 \e{1236}-96 \e{12356}$,

5) $(H_{\bar{4}} (H_{\bar{4}}
H_{\bar{4}})_{\bar{1},\bar{4},\bar{5}})_{\bar{4}}=-2487-3316
\e{1}-13264 \e{3}-646 \e{5}+19704 \e{1236}-1536 \e{1256}+384
\e{2356}+864 \e{12356}$,

 6) $\frac{2}{3} H (H_{\bar{4}}
(H_{\bar{4}} H_{\bar{4}})_{\bar{1},\bar{4},\bar{5}})_{\bar{4}}=2/3
(678025-13824 \e{5}-1152 \e{1236}-9216 \e{1256}+2304 \e{2356}-1728
\e{12356})$.}

From  3) and 5) we get that determinant norm of $\m{A}$ is equal to
$678025$. Doing calculations in a similar way we find the
inverse:\\
 {\small$\m{A}^{-1}=\frac{1}{678025}\bigl(\frac{1}{3} (44766-9765
\e{1}-95588 \e{3}+1841 \e{15}+8412 \e{26}-5176 \e{35}-71355
\e{126}-12112 \e{135}+20568 \e{236}-1554 \e{1256}+19608
\e{2356}-7488 \e{12356})+\frac{2}{3} (44766-9765 \e{1}-95588
\e{3}+1841 \e{15}+8412 \e{26}+8 \e{35}-71355 \e{126}-12112
\e{135}+18840 \e{236}-8466 \e{1256}+21336 \e{2356}-4032
\e{12356})\bigr)=\\ \frac{1}{678025}(44766-9765 \e{1}-95588
\e{3}+1841 \e{15}+8412 \e{26}-1720 \e{35}-71355 \e{126}-12112
\e{135}+19416 \e{236}-6162 \e{1256}+20760 \e{2356}-5184
\e{12356})$}.

If different formula from Table~\ref{dim6F} is employed in finding
the inverse, for example, using the first line $\frac{1}{3} H (H
(H H_{\bar{1},\bar{5}})_{\bar{4}})_{\bar{1},\bar{5}}+\frac{2}{3} H
(H_{\bar{1},\bar{4}} (H_{\bar{1},\bar{4}}
H_{\bar{4},\bar{5}})_{\bar{4}})_{\bar{4},\bar{5}}$, then
intermediate results will be different. In particular, for the
first term $\frac{1}{3} H (H (H
H_{\bar{1},\bar{5}})_{\bar{4}})_{\bar{1},\bar{5}}$ we
 find $678025/3$, and for the second term
$\frac{2}{3} H (H_{\bar{1},\bar{4}} (H_{\bar{1},\bar{4}}\\
H_{\bar{4},\bar{5}})_{\bar{4}})_{\bar{4},\bar{5}}$ we find
$1356050/3$. Thus, in this case no cancellation between nonzero
grades of both terms in the pair occurs.

\vspace{5mm}
 \textit{Example~5}. The example shows that, in principle,
one can use $n=6$ formula to find  the determinant norm and
inverse MVs of all smaller algebras, $n<6$, as well. At
first sight this may be a bit unexpected. The following example
explains how does it happen. Let's take $\cl{2}{2}$ MV,
$\m{A}=45+55 \e{1}+84 \e{12}+39 \e{134}+93 \e{234}+15 \e{1234}$,
and use the determinant norm formula of $n=6$ instead of $n=4$:
$\m{A}\m{G}=\frac{1}{3} H H (H
H)_{\bar{1},\bar{4},\bar{5}}+\frac{2}{3} H (H_{\bar{4}}
(H_{\bar{4}} H_{\bar{4}})_{\bar{1},\bar{4},\bar{5}})_{\bar{4}}$,
where $H=\m{A}\m{A}_{\bar{2},\bar{3},\bar{6}}$.  The negation of
absent grades, of course, should be ignored.

Computation steps are:

{\small 1) $H= 22501+7740 \e{1}-10410 \e{2}-8880 \e{1234}$,

2)$H H = 753425101+348315480 \e{1}-468470820 \e{2}-399617760 \e{1234}$,

3)$\frac{1}{3} H H (H H)_{\bar{1},\bar{4},\bar{5}}=67166445910339801/3$,

4) $H_{\bar{4}} H_{\bar{4}}=753425101+348315480 \e{1}-468470820 \e{2}+399617760 \e{1234}$,

5) $\frac{2}{3}(H_{\bar{4}} H_{\bar{4}})_{\bar{1},\bar{4},\bar{5}}=2*67166445910339801/3$,

6) $\m{A}\m{G}=67166445910339801=(259164901)^2$,}

\noindent where $\sqrt{\m{AG}}=259164901$ is exactly the
determinant norm of the MV calculated by $n=4$ formula. So,
calculation using the $n=6$ norm formula yields the squared
determinant of matrix representation of $\cl{2}{2}$ multivector.
It is easy then to check that the $n=6$ formula yields the inverse
of $\cl{2}{2}$ MV as well. Indeed,

{\small
7) $H (H H)_{\bar{1},\bar{4},\bar{5}}+2 (H_{\bar{4}}
(H_{\bar{4}} H_{\bar{4}})_{\bar{1},\bar{4},\bar{5}})_{\bar{4}}=
17494408312203-6017809001220 \e{1}+8093719858230 \e{2}+6904152962640
\e{1234}$

8) Finally, the inverse of $\cl{2}{2}$ MV calculated using
$n=6$ formula, is
\[\begin{split}
  \m{A}^{-1} =&\frac{1}{3*67166445910339801}\bigl(559831173421635-630567641500575 \e{1}\\
          &+ 127983403060830\e{2} - 1024375706022402 \e{12}+61927453093950 \e{34}\\
       &- 560876126302467
\e{134} - 1156984425071379 \e{234} - 302208303582585 \e{1234}\bigr)\\
=&
\frac{1}{259164901}
\bigl(
720045-811025 \e{1}+164610 \e{2}-1317534 \e{12}+79650 \e{34}\\
&-721389 \e{134}-1488093 \e{234}-388695 \e{1234}\bigr)
\end{split} \]
}
\noindent which does coincide with the inverse MV of
$\cl{2}{2}$ computed using the inverse formula for $n=4$.

And this  is not a coincidence. Explicit symbolic computations
confirm that the $n=6$ determinant norm formula when is applied to
the MV of  any algebra with vector space dimension $n<6$ indeed
yields either the determinant of matrix representation of MV (for
$n=5$), or the determinant raised in power of~2 for algebras with
the vector space dimension $n=3$ and $n=4$,  or in power of~4 for
$n=2$ and $n=1$. Exactly the same happens, when formula of
determinant norm for $n=5$ is applied to $n=4,3,2$ and $n=1$
cases. Similarly, when the determinant norm formula for $n=4$ is
used to $n=3$ MV, we obtain the determinant of matrix
representation of the MV, whereas the same procedure for $n=2$ and
$n=1$ yields the determinant raised in power of 2, etc. So, the
determinant norms disclose very interesting onion-like structure,
when each higher dimensional formula embraces all lower
dimensional ones. We think that this property may be important in
further investigations.

\section{Conclusions\label{sec:6}}

From this computer-aided study we conjecture that the inverse of a
general MV of arbitrary Clifford algebra can be always  expressed
as a linear combination of a proper number of specially
constructed self-negated products, where the initial MV may stay
in the left-most or right-most position. For the first time we
have explicitly derived  compact formulas for inverse MV that
consist of sums of two grade-negated products, Tables~\ref{invall}
and \ref{dim6F}, for all Clifford algebras of dimension $n=6$. The
formulas are independent of a particular algebra signature $(p,q)$
and may be useful in numerical and especially in symbolic
programming.  They reduce to known expressions of inverses for
vectors, bivectors etc, or their combinations.

We have also presented different formulas for inverse of even
Clifford subalgebras  for dimensions $n\le 6$
(Table~\ref{inveven}), which are important for spinor algebras. A
number of previously unknown explicit expressions were provided
for $n=5$ dimensional algebras, Table~\ref{dim5F}, some of which
may be interesting from computational point of view as well. We
believe, that formulas in Table~\ref{dim5F} exhaust all possible
single term forms of writing determinant for $n=5$ algebras using
geometric product and involutions only. We have shown that
determinant norm formulas for $n=6$ split  into two classes,
Tables~\ref{dim6F} and \ref{treetermformula}. We also presented
inverses of $n=6$ in a more symmetric three term linear
combination with all weights equal to $1/3$
(Table~\ref{treetermformula}). Though this form is not preferred
from computational point of view it may be interesting when
searching similar formulas for higher dimension Clifford algebras.
The computation of inverse MV by formulas in Table~\ref{invall}
requires considerably smaller number of multiplications since many
pieces in the formulas are iterated many times. The speed of
calculation can be improved further if MV multiplication algorithm
is realized in such a way, that it can skip calculation of some
specific grades, because we know in advance that some of grades
should vanish from the product. In this case a large number of
multiplications can be avoided.

The number $m$ of multipliers in a self-negated product of the
determinant norm is determined by total degree of polynomial of
representation of matrix in  8-periodicity
table~\cite{Lounesto97}. For algebra with $n=2$ the determinant
norm consists just of two multipliers, whereas for $n=3$ as well
as for $n=4$ one has to make products of four MV multipliers in
accord with dimension of a respective matrix in 8-periodicity
table. The Clifford algebras of vector space dimensions $5$
or $6$, require eight multivectors to construct the determinant
norm. Then, looking at $8$-periodicity table one may expect that
the determinant norm of algebras of vector space dimension $7$ and $8$ can be
constructed by multilinear combinations from products of $16$
multivectors, etc.

From the considerations above it should be  clear that the
complexity of finding explicit expression of general MV inverse
grows rapidly with the increase of algebra dimension. For example,
a direct head-on symbolic multiplication of eight general
arbitrary multivectors in $n=6$ dimensional vector space in the
worst case requires
$(2^6)^8=281\mspace{2mu}474\mspace{3mu}976\mspace{3mu}710\mspace{4mu}656$
geometric multiplications of base elements, which fortunately can
be bypassed due to significant grade reduction in  the determinant
formula as discussed in this article. However, in practice it does
not help much when searching for correct shape of determinant norm
formula. This is why reduced (shorter) MVs are always worth to try
first which, under good choice, may greatly reduce the number of
candidates for possible linear combinations. It should be stressed
that the determinant norms and inverse multivectors obtained in
the paper make computations self-contained within the geometric
product and negation structure without any need to resort to
matrix representations.

Finally,  we demonstrated that $n=6$ formula {\it alone} can
generate MV determinant norms and respective inverses of
all algebras of vector space dimension $n\le 6$ and of any
signature, \textit{albeit} using more geometric multiplications.
The identified onion-like structure of determinant norms
and  numerous shapes of inverse multivectors presented in this
article incline us to believe that similar formulas (which include
only operations of grade negation, geometric product and
summation) may exist for Clifford algebras of all
dimensions.


\end{document}